\documentclass[preprint,11pt]{aastex}
\usepackage[dvips]{epsfig}

\newcommand{\kms}{~km~s$^{-1}$} 
\newcommand{\teff}{$T_{\rm eff}$}
\newcommand{\logg}{$\log g$}

\newcommand{\vt}{$v_{\rm micro}$}

\newcommand{\BS}{BS~16545--089} 
\newcommand{\CSa}{CS~22948--093} 
\newcommand{\CSb}{CS~22965--054} 
\newcommand{\Ga}{G~64--12} 
\newcommand{\Gb}{G~64--37} 
\newcommand{\SDSSa}{SDSS~0040+16} 
\newcommand{\SDSSb}{SDSS~1033+40} 
\newcommand{\HE}{HE~1148--0037}
\newcommand{\BD}{BD+26$^{\circ}$3578}
\newcommand{\HD}{HD~84937}
\newcommand{\CD}{CD~$-24^{\circ}17504$}

\shorttitle{Lithium Abundances of Extremely Metal-Poor Turn-off Stars}
\shortauthors{Aoki et al.}

\begin{document}

\title{Lithium Abundances of Extremely Metal-Poor Turn-off
Stars\footnote{Based on data collected at the Subaru Telescope, which
is operated by the National Astronomical Observatory of Japan.}}

\author{Wako Aoki\altaffilmark{2,3}, Paul S. Barklem\altaffilmark{4}, Timothy C. Beers\altaffilmark{5}, Norbert Christlieb\altaffilmark{6}, Susumu Inoue\altaffilmark{2,7}, Ana E. Garc\'{i}a P\'{e}rez\altaffilmark{8}, John E. Norris\altaffilmark{9}, Daniela Carollo\altaffilmark{9,10}}


\altaffiltext{2}{National Astronomical Observatory, Mitaka, Tokyo,
181-8588 Japan; email: aoki.wako@nao.ac.jp}
\altaffiltext{3}{Department of Astronomical Science, The Graduate
  University of Advanced Studies, Mitaka, Tokyo, 181-8588 Japan}
\altaffiltext{4}{Department of Physics and Astronomy, Uppsala
University, Box 515, 751-20 Uppsala, Sweden; email: 
Paul.Barklem@physics.uu.se}
\altaffiltext{5}{Department of Physics and Astronomy, CSCE: Center for
  the Study of Cosmic Evolution, and JINA: Joint Institute for Nuclear
  Astrophysics, Michigan State University, East Lansing, MI
  48824-1116; email: beers@pa.msu.edu}
\altaffiltext{6}{University of Heidelberg, ZAH,
Landessternwarte K\"{o}nigstuhl 12, D-69117 Heidelberg, Germany,
N.Christlieb@lsw.uni-heidelberg.de}
\altaffiltext{7}{Present address: Department of Physics, Kyoto University, Oiwake-cho, Kitashirakawa, Sakyo-ku, Kyoto 606-8502, Japan; email:inoue@tap.scphys.kyoto-u.ac.jp}
\altaffiltext{8}{Centre for Astrophysics Research,
  STRI and School of Physics, Astronomy and Mathematics, University of
  Hertfordshire, College Lane, Hatfield AL10 9AB, United Kingdom;
  email: a.e.garcia-perez@herts.ac.uk}
\altaffiltext{9}{Research School of Astronomy and Astrophysics, The
Australian National University, Mount Stromlo Observatory, Cotter
Road, Weston, ACT 2611, Australia; email: jen@mso.anu.edu.au,
carollo@mso.anu.edu.au}
\altaffiltext{10}{INAF-Osservatorio Astronomico di Torino, Italy}

\begin{abstract} 

  We have determined Li abundances for eleven metal-poor turn-off
  stars, among which eight have [Fe/H] $<-3$, based on LTE analyses of
  high-resolution spectra obtained with the HDS on the Subaru
  telescope.  The Li abundances for four of these eight stars are
  determined for the first time by this study. Effective
    temperatures are determined by a profile analysis of
    H$\alpha$ and H$\beta$. While seven stars have Li
  abundances as high as the Spite Plateau value, the remaining 
   four objects with [Fe/H] $<-3$ have $A$(Li) $=\log$(Li/H)$ +12
  \lesssim 2.0$, confirming the existence of extremely metal-poor
  turn-off stars having low Li abundances, as reported by previous
  work. The average of the Li abundances for stars with [Fe/H]$ < -3$
  is lower by 0.2~dex than that of the stars with higher
  metallicity. No clear constraint on the metallicity dependence or
  scatter of the Li abundances is derived from our measurements
    for the stars with [Fe/H]$<-3$. Correlations of the Li abundance
  with effective temperatures, with abundances of Na, Mg and Sr, and
  with the kinematical properties are investigated, but no clear
  correlation is seen in the extremely metal-poor star sample.

\end{abstract}

\section{Introduction}\label{sec:intro}

Over the course of the past few decades, Li abundances have been
measured for many metal-poor main-sequence turn-off stars, in hopes of
placing constraints on the nature of Big Bang nucleosynthesis (BBN).
The first such investigation, by \citet{spite82}, revealed that warm
metal-poor main-sequence stars exhibited a constant Li abundance. This
so-called Spite Plateau was interpreted as a result of the synthesis
of light nuclei in the first several minutes of the evolution of the
Universe (it is conventionally assumed that Li is not destroyed at the
surface of such stars because of their shallow surface convection
zones). Since then, measurements of Li abundances have been obtained
over wide ranges of effective temperature and metallicity, confirming
that the scatter of the Li abundances in most stars with
{\teff}$>5700$~K and [Fe/H]$<-1.5$ is small, if present at all
\citep[][ and references therein]{ryan99} \footnote{[A/B] =
$\log(N_{\rm A}/N_{\rm B}) -\log(N_{\rm A}/N_{\rm B})_{\odot}$, and
$\log\epsilon_{\rm A} =\log(N_{\rm A}/N_{\rm H})+12$ for elements A
and B. Lithium abundances are conventionally presented as $A$(Li)
instead of $\log\epsilon_{\rm Li}$.}.

However, it has been recognized that the Spite Plateau value of A(Li)
($\sim 2.2$) is 0.3--0.4~dex lower than the Li abundance predicted by
standard BBN models \citep{coc04}, adopting the baryon density
determined by recent measurements of the cosmic microwave background
(CMB) radiation with the WMAP satellite \citep{spergel03}. This
discrepancy indicates the existence of poorly understood processes
that deplete Li in metal-poor turn-off stars \citep[e.g.
][]{korn06,korn07}, astration of Li in the gas that formed these stars
(perhaps by massive zero-metallicity progenitors; Piau et al. 2006),
problems in the measured abundances of Li in metal-poor stars,
systematic uncertainties in the standard BBN model predictions, or
exotic processes in the early universe arising from nonstandard
particle physics (Cyburt et al. 2008 and references therein).

Recent measurements of very metal-poor turn-off stars also suggest a
decreasing trend of Li abundances with decreasing metallicity
\citep{ryan96, ryan99, boesgaard05,asplund06}. In particular,
\citet{bonifacio07} investigated the Li abundances for a sample that
includes eight stars with [Fe/H] $ < -3$. They showed that the Li
abundances of these stars are lower than those of stars with higher
metallicity, investigating very carefully the scatter and trend of Li
abundances with metallicity and effective temperature.

Another unsolved problem is the low Li abundance found in
HE~1327--2326, a hyper metal-poor turn-off star having [Fe/H]$ < -5$
\citep{frebel05, aoki06}. Only an upper-limit on the Li abundance has
been determined, $A$(Li)$ < +0.62$ \citep{frebel08}. Although the star
has probably evolved to the subgiant branch \citep{korn09}, the
effective temperature of this object (6180~K; Frebel et al.  2005) is
still sufficiently high that depletion of Li by surface convection is
not expected. Another possible explanation for the apparent depletion
of Li is mass accretion from an evolved companion star \citep{ryan02},
in particular because this star is highly carbon-enhanced ([C/Fe]
$\sim +4.0$; Aoki et al.  2006). However, no signature of binarity has
been found yet for this object \citep{frebel08}.

Thus, concerning the $^{7}$Li abundances in very metal-poor stars, we
are confronted with three problems: (1) The discrepancy between the
observed Spite Plateau value and the prediction of standard BBN models
adopting the baryon density determined by CMB measurements; (2) A
possible trend of the Li abundance as a function of metallicity in
extremely metal-poor stars; and (3) The low Li abundance in
HE~1327--2326 (see also the summary by Piau et al. 2006). Although
possible connections between the above three problems are still
unknown, we have obtained measurements of Li abundances for several
extremely metal-poor turn-off stars in a search for hints to solving
these Li puzzles.

In this paper we report measurements of Li abundances for very and
extremely metal-poor stars. Our sample includes eight stars with
[Fe/H] $ < -3$, among which four stars are studied for the first
time. The sample selection and high-resolution spectroscopy are
described in \S~\ref{sec:obs}. Section \ref{sec:ana} reports the
determination of stellar parameters and details of the measurement of
Li abundances. Uncertainties and comparisons with previous work are
also discussed in this section. We discuss the implications of our
measurements in \S~\ref{sec:disc}, and consider possible correlations
with the derived stellar atmospheric parameters, other elemental
abundances, and the kinematics of the sample.

\section{Observations}\label{sec:obs}

High-resolution spectra of very and extremely metal-poor main-sequence
turn-off stars were obtained in the course of three different
observing programs, using the Subaru Telescope High Dispersion
Spectrograph (HDS; Noguchi et al. 2002).  Table 1 lists the objects and
details of the observations. The spectra of the first five objects
were obtained in an observing program for extremely metal-poor stars
in 2005 \citep{aoki06}. The two objects from the SDSS sample were
observed in a program for Carbon-Enhanced Metal-Poor (CEMP) stars
\citep{aoki08}. Although these two stars were selected as candidate
CEMP turn-off stars having [Fe/H] $< -3$, they turned out to show no
clear carbon excess in our high-resolution spectroscopy, thus were
good targets for studying the Li abundances in the extremely low
metallicity range. The other four bright stars were observed with very
high signal-to-noise (S/N) ratios in order to measure Li isotope
ratios (P.I. S. Inoue).

Lithium abundances for seven objects in our sample have been recently
measured with high-resolution spectroscopy by other authors. The stars
{\CSa} and {\CSb} were studied by \citet{bonifacio07}. The five bright
objects are well-known stars for Li studies. {\BD} was recently
studied by \citet{asplund06}. Li abundances of {\Ga} and {\Gb} were
measured by \citet{ryan99} and \citet{boesgaard05}. \citet{ryan99}
also determined Li abundances for {\HD}, {\BD} and {\CD}. Duplications
of targets with previous studies provide the opportunity to examine
the consistency between independent abundance measurements. The Li
abundances of the other four stars ({\BS}, {\HE}, {\SDSSa} and
{\SDSSb}) are reported for the first time by the present study.

The spectral resolution for the last four stars in Table 1 is
$R=90,000$ or $100,000$. The other objects were observed with
$R=60,000$ with 2 $\times$ 2 CCD on-chip binning. An exception is
{\SDSSb}, which was observed with $R=45,000$ to collect sufficient
photons, using a wider slit, under relatively poor seeing
conditions. The spectra cover 4100--6800~{\AA}, although the coverage
is slightly different between the individual observing programs.

\begin{figure}
\includegraphics[width=12cm]{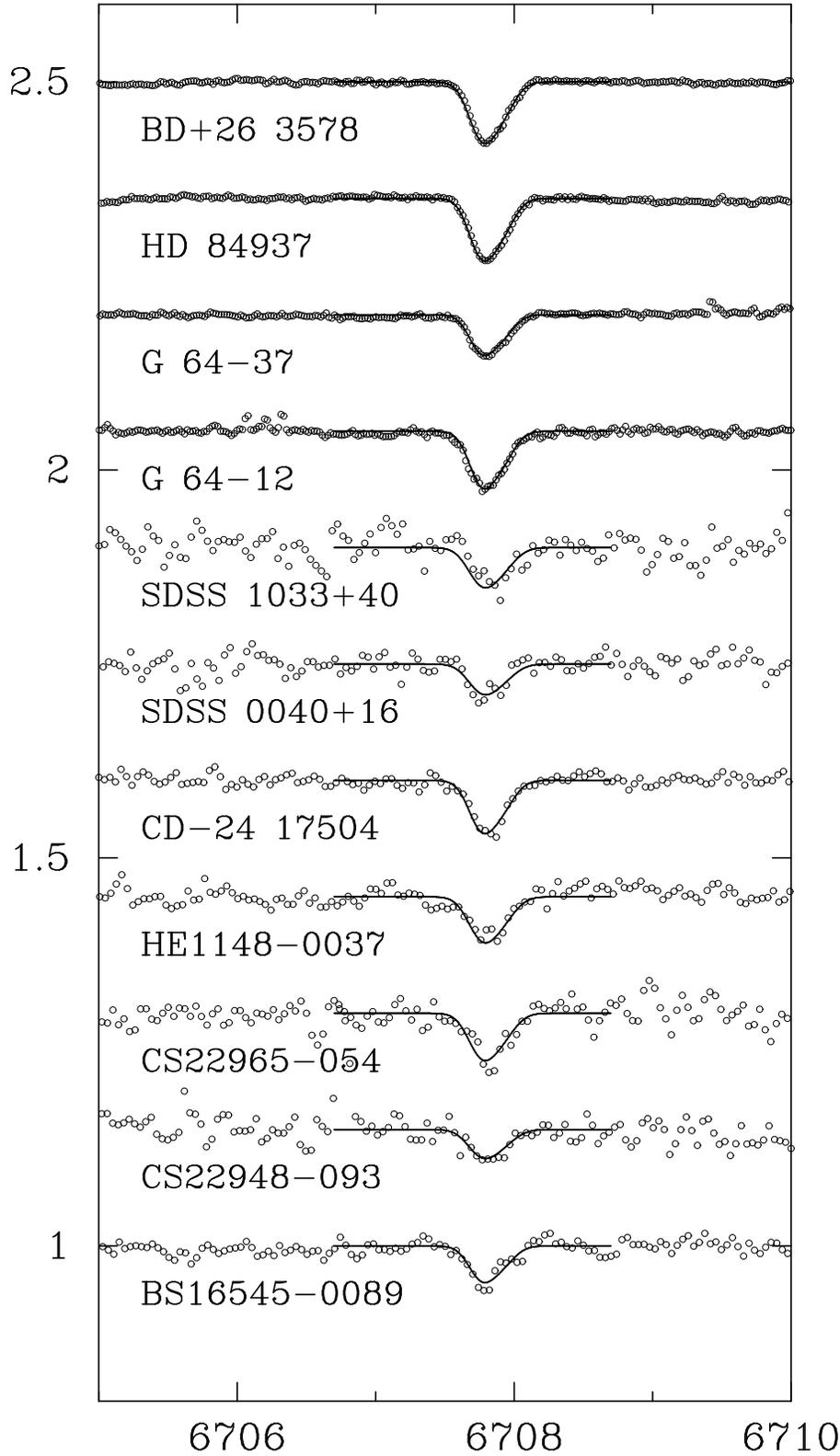}
\caption[]{
Spectra around the \ion{Li}{1} resonance line. Open circles show the
observed spectra normalized to the continuum level, which has been
vertically shifted for presentation purposes. A synthetic spectrum for
the derived Li abundance is over-plotted by a line for each star.}
\label{fig:sp}
\end{figure}

Data reduction was carried out with standard procedures using
IRAF\footnote{IRAF is distributed by the National Optical Astronomy
Observatories, which is operated by the Association of Universities
for Research in Astronomy, Inc.  under cooperative agreement with the
National Science Foundation.}. Photon counts at 6700~{\AA} are listed
in Table~\ref{tab:obs}. Spectra of this wavelength region are shown in
Figure~\ref{fig:sp}. Equivalent widths of Fe, Na, Mg, and Sr lines
were measured by fitting Gaussian profiles, in order to determine
atmospheric parameters and chemical compositions. Since the
\ion{Li}{1} $\lambda$6708 line consists of a doublet, Gaussian fitting
is not appropriate for the measurement of equivalent widths. Instead,
we determined the Li abundances by fitting synthetic spectra (\S
\ref{sec:liana}). For comparisons of equivalent widths with previous
studies, we measured the equivalent widths of this line from the
synthetic spectra (\S \ref{sec:ana}) that provided the best fits to
the observed ones. The equivalent widths of the interstellar
\ion{Na}{1} D lines, which can be used to estimate the interstellar
reddening (\S \ref{sec:ana}), were measured by direct integration of
the absorption features.


The heliocentric radial velocities listed in Table~\ref{tab:obs} were
measured from the same clean \ion{Fe}{1} lines that were used to
determine iron abundances. The uncertainty given in
Table~\ref{tab:obs} is $\sigma_{\rm v}/N$, where $\sigma_{\rm v}$ is
the standard deviation of the measurements and $N$ is the number of
lines used. In addition, one should take into account possible small
systematic errors due to the instability of the instrument, which are
at most 0.4~km~s$^{-1}$ \citep{aoki05}.

\section{Stellar Parameters and Li Abundance Measurements}\label{sec:ana}

\subsection{Effective Temperatures}\label{sec:teff}

The effective temperatures ({\teff}'s) of our program stars were determined from
profile fits to hydrogen Balmer lines, a technique used as well by \citet{asplund06}
and \citet{bonifacio07}.  We employed the H$\alpha$  and H$\beta$
lines [{\teff}(H$\alpha$) and {\teff}(H$\beta$)]. Although H$\gamma$ is
also covered by our observations, it lies at the edge of the CCD, making
continuum rectification difficult, and thus has not been used. 
The method for
continuum rectification of the observed spectra and analysis used
follows exactly that described in \citet{barklem02}, which may be
consulted for details. The most important aspects of the analysis are
that the synthetic profiles are computed assuming LTE line formation
using 1D LTE plane-parallel MARCS models \citep{asplund97}, with
convection described by mixing length theory with parameters
$\alpha=0.5$ and $y=0.5$. The most important line-broadening
mechanisms for the wings are Stark broadening and self broadening,
which are described by calculations of \citet{stehle99} and
\citet{barklem00}, respectively. The fitting is accomplished by
minimisation of the $\chi^2$ statistic, comparing the observed and
synthetic profiles. Essentially the same method was used by
\citet{asplund06}, although with some differences in the rectification
and profile comparison techniques.

We note that the assumption of LTE for formation of the Balmer
  line wings in cool stars has recently been shown to be questionable
  on the basis of theoretical non-LTE calculations \citep{barklem07},
  the role of hydrogen collisions being a major uncertainty.  Those
  calculations admit that the temperatures from LTE Balmer line wings
  could be systematically too cool by of order 100~K if hydrogen
  collisions are inefficient, although LTE is not ruled out.  Due to
  the fact that there is no strong evidence favoring any particular
  hydrogen collision model, we chose to calculate in LTE as this
  temperature scale is well studied, and it is computationally most
  practical.  However, we emphasize that LTE is not a safe middle
  ground, and will lead to temperatures systematically too cool should
  departures from LTE exist in reality.

Often, H$\alpha$ is given higher weight than higher series lines such
as H$\beta$ for reasons discussed by \citet{fuhrmann93}.  However, in
extremely metal-poor turn-off stars this is no longer obvious since
blending by metal lines becomes unimportant, and H$\beta$ becomes in
fact almost insensitive to gravity, while H$\alpha$ is quite gravity
sensitive (see table 4 of \citet{barklem02}).  Further, the
calculations by \citet{barklem07} suggest that non-LTE
effects, if they exist, will be largest in H$\alpha$.  Thus, in
combining the temperatures from H$\alpha$ and H$ \beta$, we have in
fact given H$\beta$ double the weight of H$\alpha$.

The derived effective temperatures are dependent on the gravity
assumed in the calculation. The analysis is iterated for the gravity,
which is determined from the analysis of the high-resolution spectrum
(\S~\ref{sec:ana}) for each object. The results and uncertainties of
the {\teff} determinations are listed in Table~\ref{tab:param}. The
uncertainties listed in the tables are estimated from the quality of
the spectrum and the fit.

For comparison purposes, we also estimated the effective temperatures
from the $(V-K)_{0}$ colors [{\teff}$(V-K)$],  using the effective
temperature scales of \citet{alonso96}, \citet{ramirez05b}, and
\citet{gh09}. For the estimates using the scale of \citet{alonso96}, we
assume [Fe/H]$=-3.0$ for stars having [Fe/H]$<-3.0$ following
\citet{ryan99}. The photometry data were collected from
\citet{beers07}, the 2MASS catalogue \citep{skrutskie06}, and the
SIMBAD database. The $V$ magnitudes of the SDSS stars are derived from
the $g$ magnitude and $g-r$ color, as in \citet{aoki08}. Interstellar
reddening is estimated from the dust maps of \citet{schlegel98} and
from the interstellar \ion{Na}{1} D line, using the scale of
\citet{munari97}; results are listed in Table~\ref{tab:teff}. We adopt
the reddening estimate obtained from the Na D line for the brightest
four stars ({\HD}, {\BD}, {\Ga} and {\Gb}) because these are nearby
stars, and the reddening might be overestimated from the dust maps.
The interstellar Na D lines blend with the stellar lines in the
spectrum of {\HE}, so the reddening value from the dust map is adopted
for this object. For the other six objects, the averages of the
reddening values derived from the two methods are adopted. The
differences between the $E(B-V)$ values based on the two methods are
generally 0.01 mag or smaller. The exception is for {\CSb}, which
exhibits the largest $E(B-V)$ among the sample, $E(B-V) = 0.09$ from
the dust map and 0.13 from the Na D line. The extinction in each band
is obtained from the reddening relation given by \citet{schlegel98}.

\begin{figure}
\includegraphics[width=12cm]{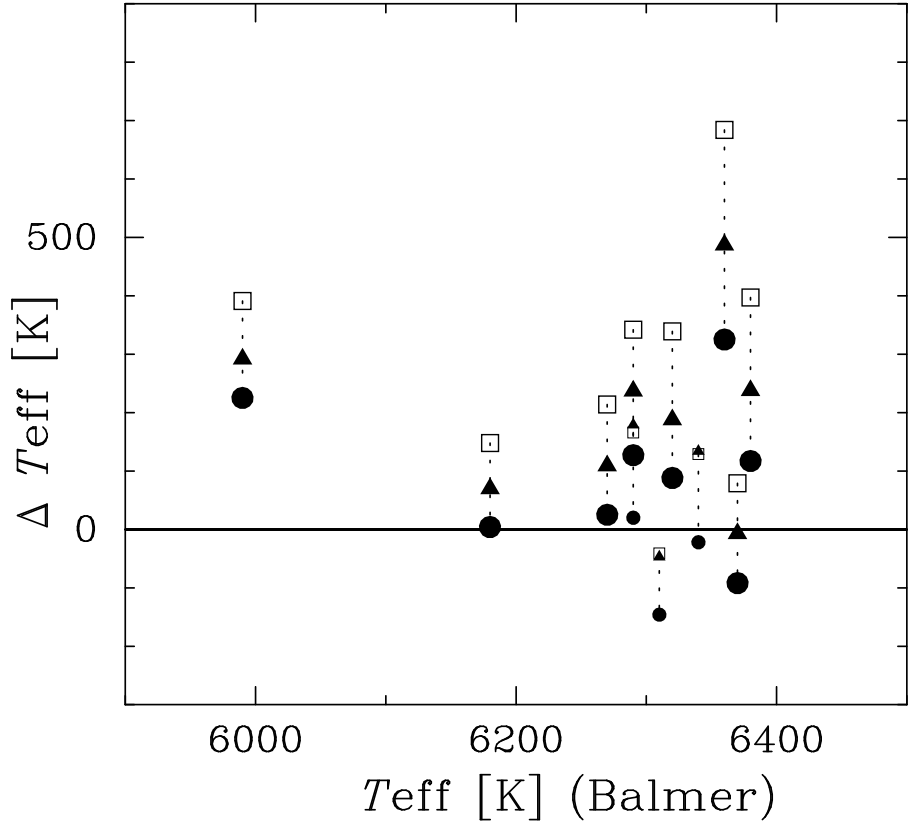}
\caption[]{Comparisons of effective temperatures derived from the
  $(V-K)_{0}$ colors with those from the Balmer line analysis. The
  differences of the effective temperature from those from the Balmer
  line analysis are shown for the scales of \citet{alonso96}
  (circles), \citet{ramirez05b} (squares), and \citet{gh09} (triangles).
}
\label{fig:teff}
\end{figure}

The photometry data and the derived effective temperatures are listed
in Table~\ref{tab:teff}. Figure~\ref{fig:teff} shows the differences
between the effective temperatures from Balmer lines and those from
$V-K$ colors for the three temperature scales. The effective
temperatures from the scales of \citet{ramirez05b} and \citet{gh09} are
systematically higher than those from \citet{alonso96}. The
differences between the results from \citet{gh09} and from
\citet{alonso96} are 100-150~K. The results from
\citet{ramirez05b} are 100-200~K higher still than those from \citet{gh09}
for stars with [Fe/H]$<-3$, while the agreement is fairly good for the
three stars having higher metallicity (shown by smaller symbols in
the figure). These differences between the effective temperature
scales were already reported by \citet{ramirez05b} and \citet{gh09}.

The comparisons with the results from Balmer line analyses indicate
that the effective temperatures from the scale of \citet{alonso96}
agree in general with those from the Balmer lines. The star showing
the largest discrepancy is {\SDSSa}. This star has a $(V-K)_{0}$
smaller than the color range ($1.1<(V-K)_{0}<1.6$) for which the
formulae of Alonso et al. are applicable. The results from the scales
of \citet{gh09} and \citet{ramirez05b} for stars with [Fe/H]$<-3$ are
systematically higher than those from Balmer lines by 150~K and 250~K,
respectively. 

Determination of the Li abundance is sensitive to the effective
temperatures adopted in the analysis.  We discuss the impact of effective
temperatures on the metallicity dependence of Li abundances in
\S~\ref{sec:disc}.

The estimate of effective temperatures for extremely metal-poor stars
from these temperature scales would be rather uncertain, because
of the small number of stars in the sample used to produce the
scales. \citet{alonso96} includes a very small number of stars for the
ranges of the color and metallicity. \citet{ramirez05a} increased the
number of such stars, which are used to produce the scale of
\citet{ramirez05b}. However, the sample is still not large (10 stars)
and includes one carbon-enhanced star and two known double-lined
spectroscopic binaries. Further measurements of effective temperatures
by the infrared flux method are desired for this metallicity range.

\subsection{Other Parameters and Abundances}\label{sec:param}

The abundance analyses were made using the grid of ATLAS9 NEWODF model
atmospheres \citep{kurucz93, castelli03}, with enhancement of the $\alpha$-elements.
Calculations of synthetic spectra for abundance analyses were made employing a
1D-LTE spectral synthesis code that is based on the same assumptions as the
model atmosphere program of \citet{tsuji78}. The surface gravities ({\logg}) are
estimated from the effective temperatures and $Y^{2}$ isochrones \citep{y2} for
[Fe/H]=$-2.5$ (for {\HD} and {\BD}) and [Fe/H] $=-3.5$ (for the others) assuming
the ages of these stars to be 12~Gyr. For the {\teff} range of our sample, two
possibilities of {\logg} exist: the subgiant case ({\logg}$\sim 3.8$) and the
main-sequence case ({\logg}$\sim 4.4$). We performed abundance analyses for
\ion{Fe}{1} and \ion{Fe}{2} (see below for details), and selected the {\logg}
that provides better agreement of iron abundances from the two species for each
star. The derived {\logg} value is not sensitive to the assumption of the age
for selecting isochrones. The results for {\logg} and the derived iron
abundances from the two species are listed in Table~\ref{tab:param}. We
re-determined effective temperature when the gravity assumed in the first
estimate of effective temperature is different from the derived one. We note
that the Fe abundance from \ion{Fe}{1} lines and the Li abundance are not
sensitive to the gravity adopted in the analysis.

The analysis, when adopting the micro-turbulent velocity ({\vt}) of 1.5
{\kms}, results in no statistically significant dependence of the iron
abundances on the strengths of the \ion{Fe}{1} lines used for the
analysis for {\BS}, {\HE}, {\Ga}, {\Gb} and {\BD}, while 1.2 {\kms} is
preferable for {\HD}. Since the number of iron lines available in the
abundance analyses of the other four stars is too small to estimate
their micro-turbulent velocity, we assume {\vt} of 1.5~{\kms} for
these objects. We note that the Li abundance derived from the weak
\ion{Li}{1} $\lambda$ 6708 line is insensitive to the micro-turbulent
velocity adopted in the analysis.

In the following discussion, we adopt the iron abundances from
\ion{Fe}{1} lines (Table~\ref{tab:param}) as the metallicity. Although
the iron abundances from \ion{Fe}{1} lines are considered to be
possibly affected by non-LTE effects \citep[][ and references
therein]{asplund05a}, our results are insensitive to the gravity
adopted in the analysis. We refer to a solar Fe abundance of $\log
\epsilon_{\odot}$(Fe)$=7.45$ \citep{asplund05b} to derive the [Fe/H]
and [X/Fe] values.

In order to investigate correlations between the abundances of Li and
other elements, Na, Mg, and Sr abundances are determined by a
  standard LTE analysis from the \ion{Na}{1} 5890 and 5896 {\AA}
lines (D lines), \ion{Mg}{1} 5172, 5183, and/or 5528~{\AA}
lines, and \ion{Sr}{2} 4078 and/or 4215 {\AA} lines,
respectively. The agreement of the abundances from two or three
  lines for each element is fairly good. The results are listed in
Table~\ref{tab:res}. Correlations between the abundances of Li and
these elements are examined in \S~\ref{sec:disc}. The Na D lines
  are known to suffer a significant NLTE effect that is dependent on
  the line strengths. The effect is, however, not large
  ($\Delta$[Na/Fe]$_{\rm NLTE}-$[Na/Fe]$_{\rm LTE}\lesssim$0.1~dex)
  for the extremely metal-poor stars that have equivalent widths of
  the D lines less than 50~m{\AA} \citep{takeda03,andrievsky07}. The
  effect could be larger ($\sim -0.3$~dex) for the less metal-poor
  stars whose equivalent widths are as large as 100~m{\AA}.

\subsection{Analyses of the Li Line}\label{sec:liana}

The Li abundances are derived from the \ion{Li}{1} 6708~{\AA} line by
fitting the observed spectra with synthetic spectra. Doppler
corrections for the observed spectra are determined by the \ion{Fe}{1}
line positions used for the above abundance measurements.
In the fitting procedure for
the four bright stars ({\HD}, {\BD}, {\Ga} and {\Gb}), a small
wavelength shift is included as a fitting parameter (see below).  The
continuum level is estimated from the wavelength range around the Li
line (6706.9--6707.4~{\AA} and 6708.2--6708.7~{\AA}). The list of the
Li lines provided by \citet{smith98} is adopted, neglecting the
contribution of $^{6}$Li.  We assume a Gaussian profile to account for
the broadening by macro-turbulence, including rotation, and by the
instrument. We assume a broadening of 7~{\kms} for the stars observed
with $R\geq 90,000$, and 8~{\kms} for the others, values that
sufficiently explain the widths of weak Fe lines. We performed the
analyses changing the value by $\pm1$~{\kms}, and confirmed that the
effect of the assumed line broadening on the derived $^{7}$Li
abundance is small ($\sim 0.01$~dex).

We then search for the Li abundance that yields the minimum
$\chi^{2}$. We found that the fit is slightly improved by allowing a
small wavelength shift for the four brightest stars, while the effect
is negligible for the others. Hence, in our procedure for fitting of
the synthetic spectra to the observations, the free parameter was the
Li abundance, and for the four brightest stars also a wavelength
shift. The number of data points to which the fitting is applied is 18
and 36 for the cases of two pixel binning and of no binning,
respectively.  The 2$\sigma$ range is adopted as the fitting error
($\sigma$[$A$(Li)]$_{\rm fit}$ in Table \ref{tab:res}). While the
fitting errors for the four bright stars ({\Ga}, {\Gb}, {\HD} and
{\BD}) are on the order of 0.02~dex and are smaller than the other
errors (see below), those for the other fainter stars are 0.06 -- 0.18
dex, depending on the S/N of the spectra.

The equivalent widths of the Li doublet are determined by integrating the
flux density of the synthetic spectra calculated for the final Li
abundance as mentioned in \S~\ref{sec:obs}. The values are listed in
Table~\ref{tab:res}.

\subsection{Uncertainties}

In addition to the above fitting errors, the uncertainties in the
adopted continuum levels, macro-turbulence, and atmospheric parameters
should be considered. The continuum level is estimated for the
wavelength range around the Li doublet, in which 25 (in the case of
two pixel binning) or 50 (in the case of no binning) data points
exist. The uncertainty is 0.1\% or smaller for the four bright stars
($S/N \gtrsim$300), while it is approximately 0.5\% for the spectra
with $S/N \sim 50$. The effects of continuum level shifts of 0.1\% and
0.5\% are estimated to be 0.01~dex and 0.04~dex, respectively, by means of
an analysis of the spectrum of {\Ga}, in which the continuum level was
changed artificially. The errors due to the uncertainty of the
macro-turbulent velocity are on the order of 0.01~dex, as mentioned
above.

The errors due to the uncertainties in atmospheric parameters are
estimated by changing the parameters by $\delta$({\teff}) $=+100$~K,
$\delta$({\logg}) $=-0.3$~dex, and $\delta$({\vt}) = +0.3~{\kms} for
{\Ga}. The effect of metallicity changes of 0.2~dex on the abundance
analyses is negligible for extremely metal-poor stars.  We
confirmed that the effects of changes of {\logg} and {\vt} on the
derived Li abundance are negligible. The Li abundance changes by
+0.07~dex when the effective temperature is changed by +100~K. Such
error estimates are scaled for the uncertainty for $T_{\rm eff}$ for
each object listed in Table~\ref{tab:param}.

The total abundance errors ($\sigma$[$A$(Li)]$_{\rm tot}$) are obtained
by adding, in quadrature, the fitting errors and the errors due to the
uncertainty of the continuum placement, the macro-turbulence, and the
effective temperature for each star (see Table~\ref{tab:res}).

\subsection{Comparisons with Previous Work}

Table~\ref{tab:comp} compares the results of the present study for
stars in common with other recent studies. The {\teff} and $A$(Li) of
{\CSa} determined by this work and by \citet{bonifacio07} agree fairly
well, while a significant discrepancy of $A$(Li) is found for
{\CSb}. This is clearly caused by the different effective temperatures
adopted in the two analyses.

The abundances of {\BD} were determined by \citet{asplund06}(where the
star is referred to as HD~338529), adopting very similar atmospheric
parameters ({\teff}, {\logg}) = (6335~K, 4.04) to those of the present
work (6340~K, 3.9). Their Fe abundance ([Fe/H] $=-2.26$ and $\log
\epsilon$(Fe) $=5.24$) from \ion{Fe}{2} lines as well as the Li
abundance agrees well with our result.

The {\teff}'s of {\Ga} and {\Gb} determined by \citet{boesgaard05} are
about 200~K lower than ours, which result in their lower $A$(Li)
values. This is, however, partially compensated by the difference of
model atmospheres used in the analysis. Boesgaard et al. adopted the
grid of \citet{kurucz93}, which assumes convective overshooting. We
found that the resulting $A$(Li) from this grid is systematically
higher by 0.06~dex than that obtained from the no overshooting models
by conducting analysis with both models. The discrepancy of $A$(Li)
derived by us and \citet{boesgaard05} is explained by the differences
of {\teff} and the model atmosphere grid. The {\teff} of
\citet{boesgaard05} is spectroscopically determined by the analysis of
\ion{Fe}{1} lines. A systematic difference between {\teff}'s from the
excitation equilibrium and those from other methods (such as
photometric colors) have been reported by previous studies
\citep[e.g., ][]{norris01, barklem05}, which might be attributed to
non-LTE effects on Fe line formation, although the recent work by
\citet{hosford08} reported no discrepancy between the {\teff} from LTE
analyses of \ion{Fe}{1} lines and {\teff} from the H$\alpha$ profile
analysis.

The {\teff}'s determined by \citet{ryan99} are systematically lower
than ours. The differences in {\teff} for {\Ga}, {\Gb} and {\CD} are
smaller than 100~K, while the differences are as large as 200~K for
{\HD} and {\BD}. The discrepancy of $A$(Li)'s between the two studies
are basically explained by the differences in the adopted effective
temperatures.

The above comparisons demonstrate that the differences of derived Li
abundances are due to differences of the effective temperature and/or
the model atmosphere grid adopted in the analysis.  These effects are
systematic, and do not essentially affect the discussion of the slope
or scatter of Li abundances if the same technique of effective
temperature estimates and similar model atmosphere grids are used.

\section{Discussion}\label{sec:disc}

\subsection{Low Li Abundances for Extremely Metal-Poor stars}

\begin{figure}
\includegraphics[width=12cm]{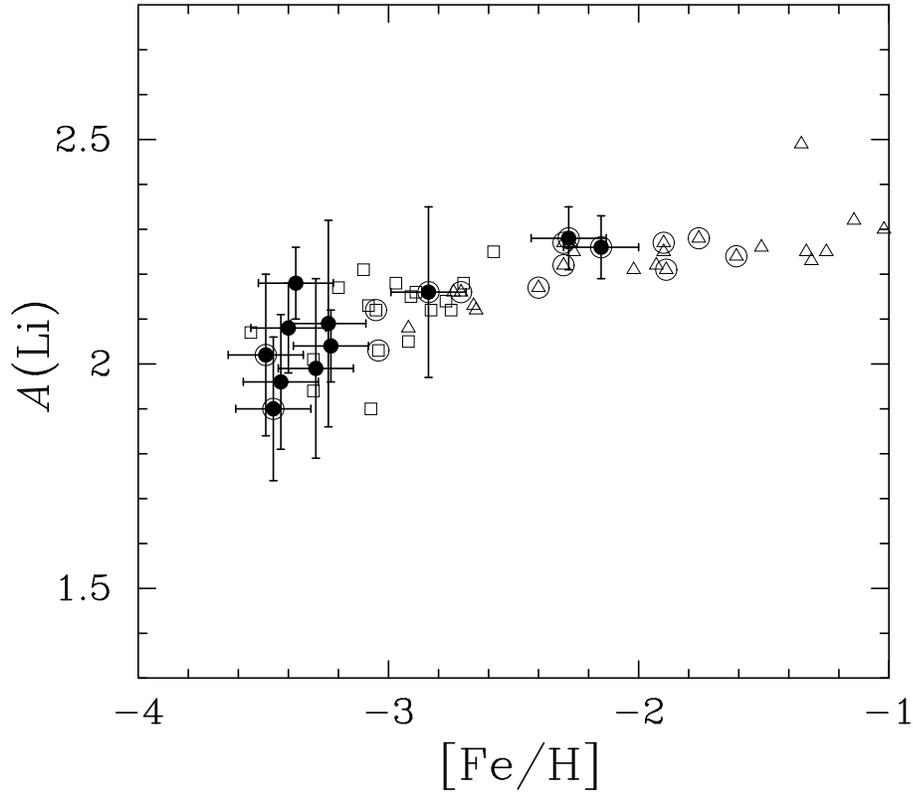}
\caption[]{ Li abundances as a function of [Fe/H]. The results of the
present work are plotted by filled circles with error bars. The open
triangles and squares indicate the results by \citet{asplund06} and
\cite{bonifacio07}, respectively. Large open circles are overplotted
for subgiant stars ({\logg} $ < 4.0$).}
\label{fig:life}
\end{figure}

Figure~\ref{fig:life} shows the Li abundances as a function of [Fe/H]
for our sample and others.\footnote{In the figure, the [Fe/H] values
of \citet{bonifacio07} and \citet{asplund06} are adopted without any
correction. The solar Fe abundance adopted by \citet{bonifacio07} is
7.51, while that of \citet{asplund06} is determined by their own
analysis of the solar spectrum. \citet{asplund06} also reported Fe
abundances from \ion{Fe}{2} lines, which are systematically higher by
0.08~dex than those from \ion{Fe}{1} lines according to the
authors. Hence, a possible shift of [Fe/H] values by at most 0.1~dex
should be taken into consideration in the comparisons of the three
works.}  We confirm that the Li abundance of the ``reference'' star
{\BD} ([Fe/H]$\sim -2.3$) determined by our analysis agrees well with
the measurement by \citet{asplund06}. By contrast, stars with [Fe/H]
$ < -3$ (the Extremely Metal-Poor, or EMP, stars) appear to have lower
Li abundances on average, and also exhibit some scatter. We discuss
this point further below.

The average of $A$(Li) (i.e., $<A$(Li)$>$) of the eight stars with
[Fe/H]$ < -3$ is 2.03, with a sample standard deviation ($\sigma$) of
0.09~dex. The standard deviation is comparable with, or slightly
smaller than, the measurement errors of our analysis
(0.07--0.23~dex). Hence, we do not detect any significant scatter of
the Li abundances in our sample of EMP stars.  The $<A$(Li)$>$ is
lower by 0.24~dex and 0.20~dex than the the average of our two reference
stars (2.27) and the average of the results of the LTE analysis by
Asplund et al. (2006) for six stars in $-2.5<$[Fe/H]$<-2.0$ (2.23),
respectively.  The difference of 0.2~dex is significant, compared to
the $\sigma$~$N^{-1/2}$=0.09/$\sqrt{8}$=0.03 dex, where $N$ is the number
of objects in the extremely metal-poor sample. We note that the Li
abundances for stars in $-2.5 < $ [Fe/H]$ < -2.0$ are well determined
by Asplund et al. (2006), and the scatter is small ($\sigma=$0.04 dex and
$\sigma$~$N^{-1/2}$=0.02 dex).  We conclude that the Li
abundances for stars with [Fe/H]$ < -3$ are 0.2~dex lower than those of stars
with higher metallicity on average, while no significant scatter or
trend with metallicity is detected in our EMP sample. A similar
conclusion was reached by \citet{bonifacio07}; our new measurements
for stars in the lowest metallicity range supports their 
results.


Such a difference can obviously be produced by a decreasing trend
(slope) of $A$(Li) as a function of metallicity. The possible slope of
$A$(Li) was discussed by \citet{bonifacio07} in detail. However, no
clear physical reason for the slope, which appears only at the lowest
metallicity range, has yet been identified.  Another possibility is
that scatter of $A$(Li) increases in the range [Fe/H] $< -3$, and, as
a result, the average decreases. This case would be relatively easily
explained by depletion of Li, although some reason for the metallicity
dependence of the depletion is also required.

If the effective temperatures estimated from the $(V-K)_{0}$
  colors using the scales of \citet{ramirez05a} and \citet{gh09} are employed, the
  Li abundances of our stars would be systematically higher. The effective
  temperatures derived using the scale of \citet{gh09} are
  systematically higher by 220~K and 150~K than the values from the
  Balmer line analysis for EMP stars and for stars with [Fe/H]$\sim -2.3$,
  respectively, resulting in 0.15~dex and 0.10~dex higher Li
  abundances. Although the difference of Li abundances between the EMP
  stars and less metal-poor stars becomes slightly smaller than the
  result derived adopting the effective temperature from the Balmer line analysis,
  the difference is still significant. By contrast, if the effective
  temperatures estimated using the scales of \citet{ramirez05a} are
  adopted, the values are 290~K and 150~K higher than those from the
  Balmer lines, resulting in 0.20~dex and 0.10~dex higher Li
  abundances. In this case, the difference of Li abundances between
  the stars with [Fe/H]$<-3$ and $>-2.5$ is only 0.1~dex, which is no longer
  significant compared to our measurement errors.

\subsection{Correlations with Stellar Parameters, Elemental Abundances, and
Kinematics}

Although no statistically significant scatter of $A$(Li) is found for
our full EMP sample, within our measurement errors, the existence of
{\it some} star-to-star differences in $A$(Li) is suggested. For
instance, even if the stars with the largest measurement errors are
excluded from the evaluation, a similar scatter of $A$(Li) remains. A
difference of 0.14~dex is found in $A$(Li) for the two bright
stars {\Ga} and {\Gb}, as found by \citet{boesgaard05} and
  \citet{nissen05}. In this subsection, we investigate correlations
between $A$(Li) and the adopted stellar parameters, in order to search
for a hint for understanding the lower Li abundances among the EMP
stars.

\begin{figure}
\includegraphics[width=12cm]{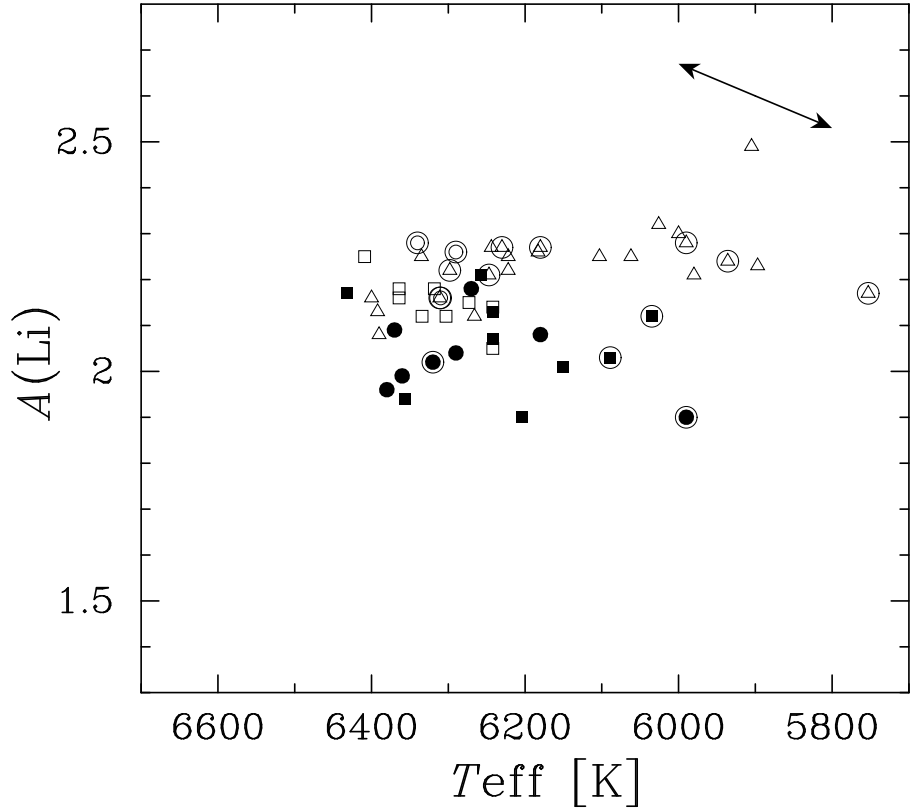}
\caption[]{
Li abundances as a function of effective temperature. The results of
the present work are plotted by circles, while the triangles and
squares indicate the results by \citet{asplund06} and
\cite{bonifacio07}, respectively. Filled symbols indicate stars having 
[Fe/H] $ \leq -3.0$. 
Large open circles are overplotted for subgiant stars. The effect of
$\delta${\teff}$=\pm 100$~K in the analysis on the Li abundance is shown by the
arrow.}
\label{fig:liteff}
\end{figure}

Figure~\ref{fig:liteff} shows $A$(Li) as a function of {\teff}. No
clear correlation can be seen in this figure. It should be noted that
the random error of the effective temperature propagates into the
derived Li abundance, which is represented by the arrows shown in the
diagram. An increasing trend of Li abundance with decreasing {\teff}
is potentially influenced by this error.

Our sample includes stars that have already evolved to the subgiant
branch.  Their effective temperatures during the main-sequence stage
should be higher than the current values, and might be as high as the
hottest stars among the main-sequence sample. The stars having $\log g
< 4.0$ are over-plotted by large circles representing candidate
subgiants in the figure. If these stars are excluded, we find that
stars cooler than 6150~K exhibit higher and almost constant Li
abundances, while the warmer stars show some scatter. However, this
probably reflects a metallicity effect, as the cooler stars (excluding
subgiants) are objects having [Fe/H] $> -2$ studied by
\citet{asplund06}, while the warmer stars have [Fe/H] $ < -2$. That
is, the sample of very metal-poor stars ([Fe/H] $ <-2$) have
{\teff}$>6150$~K, or are subgiants. This could be due to a bias in the
sample selection caused by the fact that distant stars must be
observed to cover lower metallicity ranges, and intrinsically faint
stars are not sampled. Other than this point, no clear correlation
appears between the Li abundances and effective temperature, even if
subgiants are removed from the plot, or if the highest temperature
among the sample ($\sim 6500$~K) is assigned for them as their {\teff}
during their main-sequence phase.

\begin{figure}
\includegraphics[width=12cm]{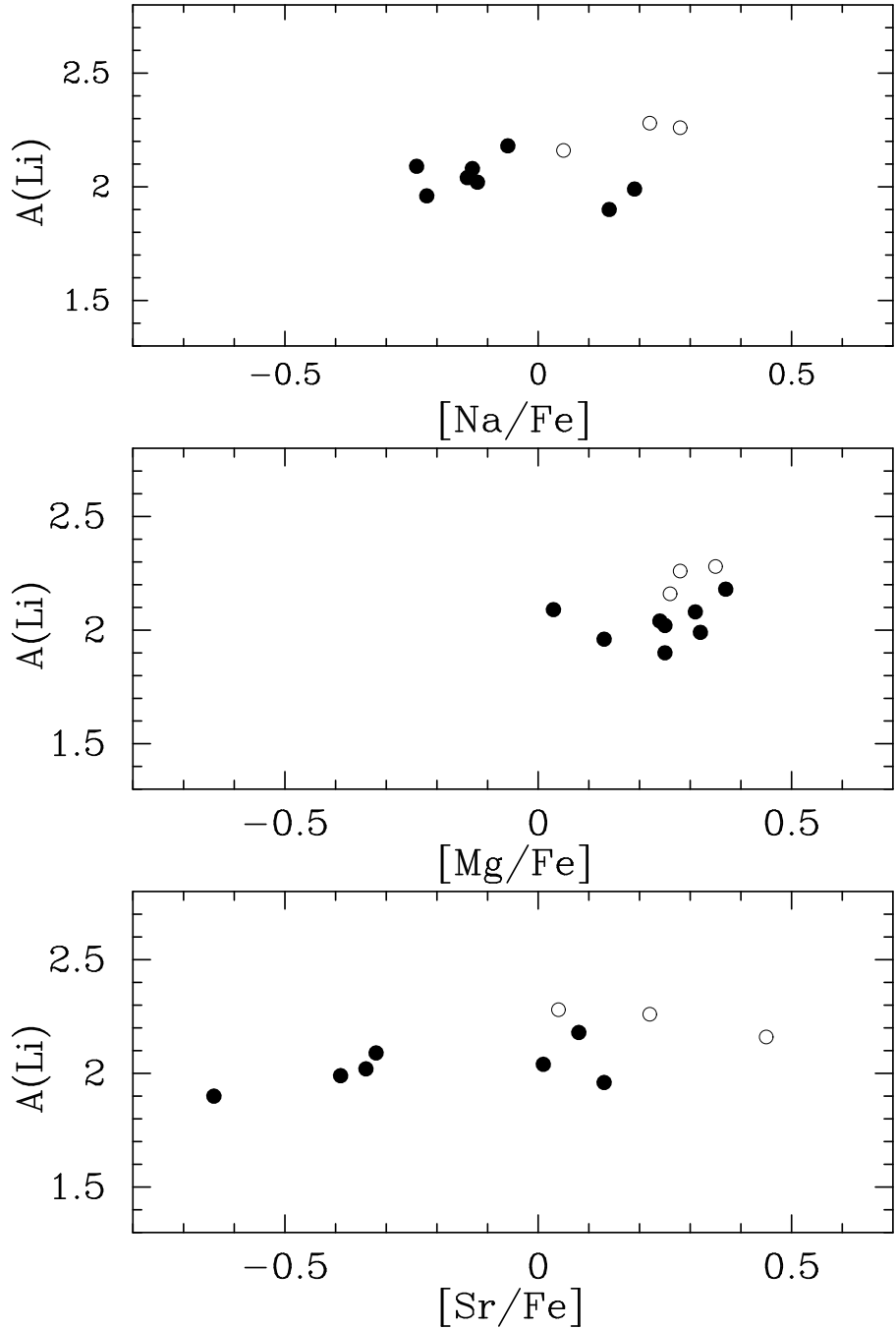}
\caption[]{Li abundances as functions of [Na/Fe], [Mg/Fe], and [Sr/Fe] for our
sample.  Filled circles indicate stars having [Fe/H] $\leq -3.0$. The
upper limit of the Sr abundance of {\CD} ([Sr/Fe] $<-0.9$) is not
plotted in this figure.}
\label{fig:limgsr}
\end{figure}

Figure~\ref{fig:limgsr} shows Li abundances as functions of [Na/Fe],
[Mg/Fe], and [Sr/Fe]. An anti-correlation between the Li abundance and
the [Na/Fe] ratio was reported for the globular cluster NGC~6752 by
\citet{pasquini05}. Such a trend is, however, not seen in our sample
(Fig.~\ref{fig:limgsr}): the two low Li stars {\HE} and {\SDSSa} have
comparatively high [Na/Fe] ratios, while that of the other low Li
star, {\CSa}, is low. If the NLTE effects on the Na abundances are
  taken into consideration, the abundances of less metal-poor stars
  become lower by about 0.3~dex. However, this correction does not
  result in any correlation between the Li and Na abundances.

No significant scatter is found for [Mg/Fe] in our sample. Also, no
correlation is found between the abundances of Li and the Mg
abundances. In contrast, a large scatter of measured Sr abundances
exists, which could reflect the contribution of a neutron-capture
process that is efficient in the very early Galaxy \citep{truran02,
travaglio04, aoki05}, as well as the contribution of the main
r-process to higher metallicity objects. No correlation is found
between the Li and Sr abundances, indicating that the Li production or
depletion is not likely to be related to neutron-capture
nucleosynthesis processes.
   
We are also interested in exploring whether the kinematics of the
stars with lower $A$(Li) exhibit any peculiarities that distinguish them
from the rest of the stars. For this exercise, we combined our present
sample with that of Bonifacio et al. (2007). Proper motions of SDSS
stars are listed in the public DR6 database, while those of other
stars are taken from the NOMAD database (Zacharias et
al. 2004). Distances were estimated from the luminosity
classifications given by either the present paper or in Bonifacio et
al. (2007), applying the methods described by Beers et
al. (2000). Stars with $\log g > 4.0$ were considered dwarfs, those
with $3.5\le \log g \le 4.0$ were considered main-sequence turnoff
stars, and those with $\log g < 3.5$ were considered subgiants. The
adopted surface gravities, metallicities, photometry, distances,
radial velocities, and proper motions are listed in
Table~\ref{tab:kine1}. These data were used to derive the full space
motions and other orbital information, following the procedures
described by Carollo et al. (2007); results are listed in
Table~\ref{tab:kine2}.

\begin{figure}
\includegraphics[width=12cm]{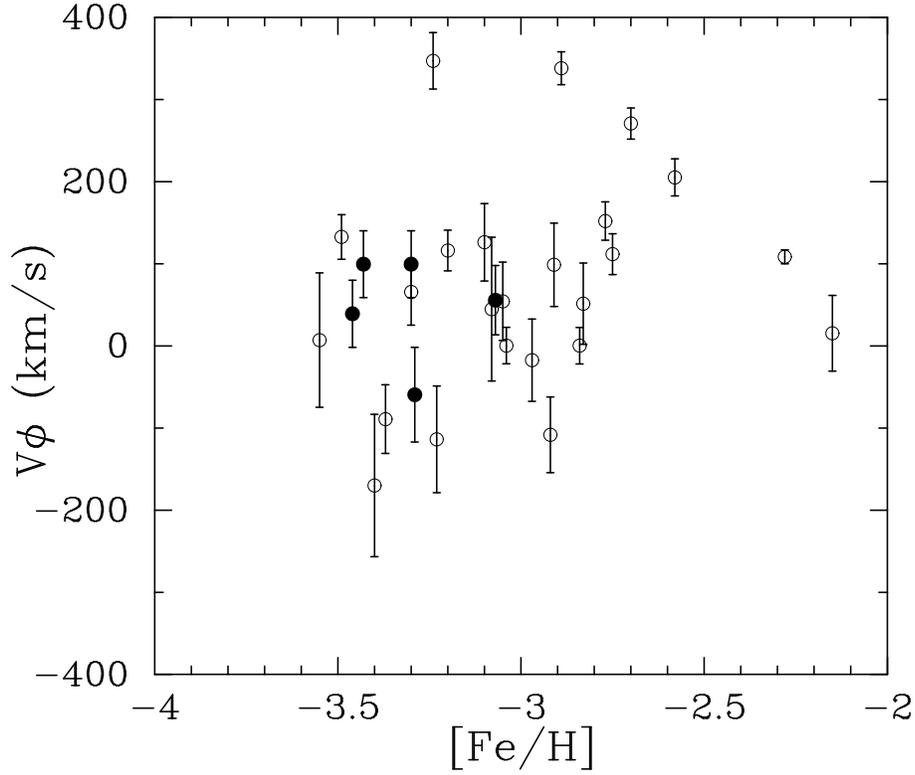} 
\caption[]{$V_{\phi}$ as a function of [Fe/H] for our sample and that
of \citet{bonifacio07}.  The filled circles indicate stars with $A$(Li)
$< 2.0$, while the open ones indicate stars with $A$(Li) $> 2.0$. A
typical error of Fe abundance determination is 0.15~dex. Note the
presence of several stars with rather high $V_{\phi}$. }
\label{fig:kine}
\end{figure}

Figure~\ref{fig:kine} shows the derived rotational velocity with
respect to the Galactic center, $V_{\phi}$, as a function of
[Fe/H]. The stars with low values of $A$(Li) are labeled by filled
circles. We note that three stars with ``normal'' $A$(Li) are on highly
prograde orbits (an additional star, CS~31061--0032, with $V_{\phi}$ =
205~{\kms}, may be a member of the metal-weak thick disk). This is
somewhat unexpected, given the essentially zero net rotation of the
inner halo, and net retrograde rotation of the outer halo. We plan to
study these stars in more detail in the near future.

Based on the derived space velocities and orbital parameters, an
attempt was made to assign approximate population memberships for
these stars, listed in the second column of Table~\ref{tab:kine2}. The
assignments take into account the values of the velocity ellipsoids
derived for the inner and outer halo, and the thick disk (including
the metal-weak thick disk; D. Carollo et al., in preparation), as well
as the derived $Z_{\rm max}$ (the maximum distance from the Galactic
plane in the vertical direction) for each star. We did not attempt to
assign a membership to the four highly prograde stars, because, as was
mentioned above, they require further investigation (FI in the second
column of Table~\ref{tab:kine2}). Note that in some cases it was not
possible to uniquely distinguish a single population assignment, so
multiple assignments are given.
 
Inspection of these assignments indicates no tendency for the stars
with low $A$(Li) to be associated preferentially with either the
inner- or outer-halo populations. This is similar to the conclusions
drawn by Bonifacio et al. (2007) based on inspection of radial
velocities alone. However, the sample size (especially of low $A$(Li)
stars) and the existing errors on the Li abundance determinations
preclude a final determination.

\subsection{Summary and Concluding Remarks}

Our measurements of the Li doublet for extremely metal-poor stars 
  based on the effective temperatures from the Balmer lines showed
that the Li abundances in the metallicity range [Fe/H] $ < -3$ are
lower, on average, than for stars with higher metallicity. The
  same conclusion is also reached by adopting the effective
  temperatures from $(V-K)_{0}$ using the scales of
  \citet{alonso96} and \citet{gh09}, although the Li abundances are
  systematically higher if the latter is adopted. If
  the temperature scale of \citet{ramirez05a} for $(V-K)_{0}$ is adopted, the
  dependence of the Li abundances on metallicity becomes marginal.

Although no scatter or trend of Li abundances is detected within the
sample of extremely metal-poor stars, the metallicity dependence of
the Li abundance could be a key to understanding the Li problems found
for metal-poor stars.  Observations to obtain better quality spectra
for such extremely metal-poor stars, to improve measurements of the Li
feature itself, as well as for improved determination of {\teff} from
the H$\alpha$ and H$\beta$  profiles, will provide a useful constraint on the
possible scenarios proposed to explain the Li problems. Moreover,
measurements of Li abundances for even lower metallicity stars ([Fe/H]
$< -3.5$) are also vital. Measurements of Li abundances in this
metallicity range have been reported so far only for one system, a
double-lined spectroscopic binary \citep[CS~22876--032: ][]{norris00,
  hernandez08}. We plan additional investigations of this metallicity
range by high-resolution spectroscopic studies of extremely metal-poor
star candidates discovered by the Hamburg/ESO survey and SDSS/SEGUE in
the near future.

\acknowledgments

The authors would like to thank the anonymous referee for useful
comments for improving this paper. W.~A. is supported by a
Grant-in-Aid for Science Research from JSPS (grant 18104003). P~.S.~B
is a Royal Swedish Academy of Sciences Research Fellow supported by a
grant from the Knut and Alice Wallenberg Foundation. P.~S.~B also
acknowledges support from the Swedish Research
Council. T.~C.~B. acknowledges support from the US National Science
Foundation under grants AST 04-06784 and AST 07-07776, as well as from
grants PHY 02-16783 and PHY 08-22648; Physics Frontier Center/Joint
Institute for Nuclear Astrophysics (JINA). N.C. acknowledges support
from the Knut and Alice Wallenberg Foundation.

\clearpage

\clearpage

\clearpage

\clearpage

\clearpage

\begin{deluxetable}{llcrcr}
\tablewidth{0pt}
\tabletypesize{\footnotesize}
\tablecaption{\label{tab:obs} PROGRAM STARS AND OBSERVATIONS}
\tablehead{
Star & Obs. date & Exp.\tablenotemark{a} & counts\tablenotemark{b} &
JD & $V_{\rm H}$ (km s$^{-1}$) \tablenotemark{c} }
\startdata
{\BS} &  28 Feb. 2005   & 225 & 16,500  & 2453429.86 & $-161.54 \pm 0.09$ \\
{\CSa} & 19 June 2005   & 174 & 5,900   & 2453541.06 & $369.24 \pm 0.30$ \\
{\CSb} & 20 June 2005   & 100 & 5,000   & 2453541.97 & $-281.67 \pm 0.20$ \\
{\HE} &  27 Feb. 2005   & 160 & 15,000  & 2453428.94 & $-10.88 \pm 0.16$ \\
{\CD} &  20 June 2005   & 20 & 14,000 & 2453542.12 & $136.61 \pm 0.13$ \\
{\SDSSa}\tablenotemark{d} & 14 Sep. 2006 & 80 & 4,800   & 2453992.90 & $-49.43 \pm 0.07$ \\
{\SDSSb}\tablenotemark{e} & 10 Feb. 2007 & 160 & 4,100   & 2454141.98 & $-132.87 \pm 0.27$ \\
{\Ga} &  22 Feb. 2003   & 300 & 103,000 & 2453507.76 & $434.68 \pm 0.03$ \\
{\Gb} &  18 May 2005    & 420 & 211,000 & 2453508.83 & $76.65 \pm 0.06$ \\
{\HD} &  22 Feb. 2003   & 180 & 626,000 & 2452692.83 & $-14.63 \pm 0.03$\\
{\BD}\tablenotemark{f} & 17 May 2005 & 132 & 342,000 & 2453508.00 & $-91.23 \pm 0.03$ \\
\enddata
\tablenotetext{a}{Exposure time (minutes)}
\tablenotetext{b}{Photon counts per pixel  at 6700~{\AA}}
\tablenotetext{c}{Heliocentric radial velocoty}
\tablenotetext{d}{SDSS J004029.17+160416.2 = 0418-51884-574} 
\tablenotetext{e}{SDSS J103301.41+400103.6 = 1430-53002-498} 
\tablenotetext{f}{= HD~338529}
\end{deluxetable}

\begin{deluxetable}{lcccccccc}
\tablewidth{0pt}
\tabletypesize{\footnotesize}
\tablecaption{\label{tab:param} ATMOSPHERIC PARAMETERS}
\tablehead{
Object & {\teff}(H$\alpha$) & {\teff}(H$\beta$) & {\teff}(adopted) & $\sigma$({\teff}) & {\logg} & {\vt} &
$A$(Fe)$_{\rm Fe I}$ & $A$(Fe)$_{\rm Fe II}$ \\
       & (K) & (K) & (K) & (K) & & ({\kms}) & &  
}
\startdata
{\BS}    & 6380 & 6290 & 6320 & 150 & 3.9 & 1.5                  & 3.96 & 3.88 \\
{\CSa}   & 6320 & 6410 & 6380 & 150 & 4.4 & 1.5\tablenotemark{a} & 4.02 & 4.02 \\
{\CSb}   & 6390 & 6270 & 6310 & 200 & 3.9 & 1.5\tablenotemark{a} & 4.61 & 4.56 \\
{\HE}    & 6100 & 5940 & 5990 & 200 & 3.7 & 1.5                  & 3.99 & 3.90 \\
{\CD}    & 6150 & 6190 & 6180 & 150 & 4.4 & 1.5                  & 4.05 & 4.17 \\
{\SDSSa} & 6350 & 6360 & 6360 & 200 & 4.4 & 1.5\tablenotemark{a} & 4.16 & 4.25 \\ 
{\SDSSb} & 6380 & 6370 & 6370 & 200 & 4.4 & 1.5\tablenotemark{a} & 4.22 & 4.29 \\
{\Ga}    & 6260 & 6280 & 6270 & 100 & 4.4 & 1.5                  & 4.08 & 4.20 \\
{\Gb}    & 6310 & 6280 & 6290 & 100 & 4.4 & 1.5                  & 4.23 & 4.33 \\
{\HD}    & 6330 & 6270 & 6290 & 100 & 3.9 & 1.2                  & 5.30 & 5.35 \\ 
{\BD}    & 6370 & 6330 & 6340 & 100 & 3.9 & 1.5                  & 5.17 & 5.20 \\
\enddata
\tablenotetext{a}{Assumed values (see text).}
\end{deluxetable}

\begin{deluxetable}{lccccccccc}
\tablewidth{0pt}
\tabletypesize{\footnotesize}
\tablecaption{\label{tab:teff} PHOTOMETRY DATA AND EFFECTIVE TEMPERATURE}
\tablehead{
Star    & $V$ &  $B-V$ & $V-K$ & $E(B-V)_{\rm SFD}$\tablenotemark{a} & $E(B-V)_{\rm Na}$\tablenotemark{a} & $E(B-V)_{\rm adopt}$\tablenotemark{a} & \multicolumn{3}{c}{{\teff}$(V-K)$\tablenotemark{b}} [K] \\
\cline{8-10}
&&&&&&& A96 & RM05 & GH09 
}
\startdata
{\BS}    & 14.450 & 0.318 & 1.248 & 0.026 & 0.020 & 0.023 & 6408 & 6659 & 6508 \\
{\CSa}   & 15.180 & 0.360 & 1.175 & 0.018 & 0.008 & 0.013 & 6497 & 6777 & 6618 \\
{\CSb}   & 15.069 & 0.497 & 1.620 & 0.089 & 0.131 & 0.110 & 6180 & 6269 & 6263 \\
{\HE}    & 13.614 & 0.404 & 1.342 & 0.022 &\nodata& 0.022 & 6215 & 6381 & 6282 \\
{\CD}    & 12.18  & 0.33  & 1.373 & 0.025 & 0.027 & 0.027 & 6184 & 6328 & 6250 \\
{\SDSSa} & 15.231 & 0.310 & 1.202 & 0.047 & 0.058 & 0.052 & 6685 & 7044 & 6847 \\
{\SDSSb} & 15.990 & 0.360 & 1.289 & 0.013 & 0.018 & 0.015 & 6278 & 6449 & 6363 \\
{\Ga}    & 11.453 & 0.385 & 1.245 & 0.028 & 0.003 & 0.003 & 6295 & 6484 & 6379 \\
{\Gb}    & 11.140 & 0.370 & 1.217 & 0.027 & 0.014 & 0.014 & 6417 & 6632 & 6527 \\
{\HD}    & 8.28   & 0.41  & 1.218 & 0.037 & 0.007 & 0.007 & 6310 & 6456 & 6469 \\
{\BD}    & 9.37   & 0.35  & 1.226 & 1.106 & 0.010 & 0.010 & 6318 & 6469 & 6474 \\
\enddata
\tablenotetext{a}{$E(B-V)$ estimated from the dust maps of
\citet{schlegel98} [$E(B-V)_{\rm SFD}$], and from the interstellar
\ion{Na}{1} D line [$E(B-V)_{\rm Na}$], as well as the adopted value
[$E(B-V)_{\rm adopt}$].}
\tablenotetext{b}{Effective temperatures estimated from $(V-K)_{0}$
using the scale of A96 \citep{alonso96}, RM05 \citep{ramirez05b}, and
GH09 \citep{gh09}.}
\end{deluxetable}

\begin{deluxetable}{lcccccccc}
\tablewidth{0pt}
\tabletypesize{\footnotesize}
\tablecaption{\label{tab:res} Li AND OTHER ELEMENTAL ABUNDANCES}
\tablehead{
Object  & [Fe/H] & $W$ (m{\AA}) & $A$(Li) & $\sigma$[$A$(Li)]$_{\rm fit}$ & $\sigma[A$(Li)]$_{\rm tot}$ & [Na/Fe] & [Mg/Fe] & [Sr/Fe] 
}
\startdata

{\BS}    & $-$3.49 & 15.3 & 2.02 & 0.12 & 0.17 & $-$0.12 & 0.25 & $-$0.34 \\
{\CSa}   & $-$3.43 & 12.4 & 1.96 & 0.12 & 0.17 & $-$0.22 & 0.13 &    0.13 \\
{\CSb}   & $-$2.84 & 20.3 & 2.16 & 0.12 & 0.19 &  0.05 & 0.26 &    0.45 \\
{\HE}    & $-$3.46 & 19.3 & 1.90 & 0.08 & 0.17 &  0.14 & 0.25 & $-$0.64 \\
{\CD}    & $-$3.40 & 21.2 & 2.08 & 0.06 & 0.13 & $-$0.13 & 0.31 & $<-$0.9 \\
{\SDSSa} & $-$3.29 & 13.6 & 1.99 & 0.14 & 0.20 &  0.19 & 0.32 & $-$0.39 \\
{\SDSSb} & $-$3.24 & 16.5 & 2.09 & 0.18 & 0.23 & $-$0.24 & 0.03 & $-$0.32 \\
{\Ga}    & $-$3.37 & 22.4 & 2.18 & 0.02 & 0.07 & $-$0.06 & 0.37 &  0.08 \\
{\Gb}    & $-$3.23 & 16.2 & 2.04 & 0.02 & 0.07 & $-$0.14 & 0.24 &  0.01 \\
{\HD}    & $-$2.15 & 24.7 & 2.26 & 0.02 & 0.07 &  0.28 & 0.28 &  0.22 \\
{\BD}    & $-$2.28 & 23.7 & 2.28 & 0.02 & 0.07 &  0.22 & 0.35 &  0.04 \\
\enddata
\end{deluxetable}

\begin{deluxetable}{lcccccccccc}
\tablewidth{0pt}
\tabletypesize{\footnotesize}
\tablecaption{\label{tab:comp} COMPARISON WITH PREVIOUS WORK}
\tablehead{
Object  & \multicolumn{2}{c}{this work} && \multicolumn{3}{c}{previous work (1)} && \multicolumn{3}{c}{previous work (2)} \\
         \cline{2-3} \cline{5-7} \cline{9-11} 
        & {\teff}(K) & $A$(Li) && {\teff}(K) & $A$(Li) & ref. && {\teff}(K)  & $A$(Li) & ref. 
}
\startdata
{\CSa}   & 6380 & 1.96 && 6356 & 1.94 & \citet{bonifacio07} && & & \\
{\CSb}   & 6310 & 2.16 && 6089 & 2.03 & \citet{bonifacio07} && & & \\
{\BD}    & 6340 & 2.28 && 6335 & 2.25 & \citet{asplund06}   && 6150 & 2.15 & \citet{ryan99} \\
{\Ga}    & 6270 & 2.18 && 6074 & 2.15 & \citet{boesgaard05} && 6222 & 2.14 & \citet{ryan99} \\
{\Gb}    & 6290 & 2.04 && 6122 & 1.97 & \citet{boesgaard05} && 6240 & 2.09 & \citet{ryan99} \\
{\CD}    & 6180 & 2.08 &&      &      &                     && 6070 & 1.97 & \citet{ryan99} \\
{\HD}    & 6290 & 2.26 &&      &      &                     && 6160 & 2.17 & \citet{ryan99} \\
\enddata
\end{deluxetable}

\begin{deluxetable}{lccrcccrrrrr}
\tablewidth{0pt}
\tabletypesize{\footnotesize}
\tablecaption{\label{tab:kine1}Surface Gravities, Metallicities, Photometry, Distance Estimates, Radial Velocities, and Proper Motions}
\tablehead{
   Star Name   & $\log g$ & [Fe/H] &    $V$  &  $B-V$ &$E(B-V)$&  Dist  &$V_{\rm r}$  &$\mu_{\alpha}$ &$\sigma_{\mu_{\alpha}}$&  $\mu_{\delta}$ & $\sigma_{\mu_{\delta}}$  \\
\cline{9-12}
               &       &        &         &        &        &  (kpc) &(km~s$^{-1}$)&  \multicolumn{4}{c}{(mas yr$^{-1}$)} 
}
\startdata
BS~16023--0046  &  4.50 &  $-$2.97 &  14.170 &  0.378 &  0.017 &  0.935 &   $-$7.5      & $-$40.0         &   2.0               &  $-$40.0          &   1.0  \\
BS~16968--0061  &  3.75 &  $-$3.05 &  13.260 &  0.430 &  0.048 &  0.712 &  $-$80.7      & $-$38.0         &  11.0               &  $-$36.0          &   8.0  \\
BS~17570--0063  &  4.75 &  $-$2.92 &  14.510 &  0.330 &  0.039 &  1.275 & $-$184.4      &  38.0         &   4.0               &  $-$30.0          &   1.0  \\
CS~22177--0009  &  4.50 &  $-$3.10 &  14.270 &  0.401 &  0.044 &  0.869 & $-$208.4      &  11.8         &   5.9               &  $-$62.5          &   5.8  \\
CS~22888--0031  &  5.00 &  $-$3.30 &  14.900 &  0.413 &  0.014 &  0.968 & $-$125.1      &  47.1         &   5.2               &  $-$20.6          &   5.2  \\
CS~22948--0093  &  4.25 &  $-$3.30 &  15.180 &  0.360 &  0.015 &  1.346 &  364.3      & $-$15.6         &   5.1               &  $-$22.8          &   4.8  \\
CS~22953--0037  &  4.25 &  $-$2.89 &  13.640 &  0.367 &  0.027 &  0.753 & $-$163.3      &  31.3         &   6.1               &   37.0          &   6.1  \\
CS~22965--0054  &  3.75 &  $-$3.04 &  15.069 &  0.497 &  0.131 &  1.532 & $-$281.5      &  26.0         &   1.0               &    0.0          &   3.0  \\
CS~22966--0011  &  4.75 &  $-$3.07 &  14.555 &  0.422 &  0.013 &  0.830 &  $-$13.5      &  15.1         &   6.0               &  $-$42.2          &   6.0  \\
CS~29499--0060  &  4.00 &  $-$2.70 &  13.030 &  0.370 &  0.019 &  0.726 &  $-$58.7      &  15.4         &   4.9               &   22.4          &   4.9  \\
CS~29506--0007  &  4.00 &  $-$2.91 &  14.180 &  0.382 &  0.045 &  1.189 &   56.4      &  $-$3.6         &   7.9               &  $-$30.3          &   7.9  \\
CS~29506--0090  &  4.25 &  $-$2.83 &  14.330 &  0.399 &  0.046 &  1.035 &  $-$21.3      & $-$12.5         &   7.9               &  $-$39.1          &   7.9  \\
CS~29518--0020  &  4.50 &  $-$2.77 &  14.003 &  0.415 &  0.023 &  0.741 &  $-$22.2      &  26.1         &   4.8               &  $-$10.4          &   4.8  \\
CS~29518--0043  &  4.25 &  $-$3.20 &  14.566 &  0.371 &  0.019 &  1.064 &  144.8      &  26.0         &   5.0               &   $-$6.0          &   1.0  \\
CS~29527--0015  &  4.00 &  $-$3.55 &  14.260 &  0.400 &  0.021 &  1.114 &   50.9      &  48.6         &  12.8               &  $-$22.5          &  12.8  \\
CS~30301--0024  &  4.00 &  $-$2.75 &  12.950 &  0.420 &  0.064 &  0.646 &  $-$67.7      &$ -$33.8         &   1.8               &  $-$23.1          &   1.7  \\
CS~30339--0069  &  4.00 &  $-$3.08 &  14.750 &  0.360 &  0.009 &  1.572 &   34.9      &  22.0         &   5.0               &  $-$16.0          &  12.0  \\
CS~31061--0032  &  4.25 &  $-$2.58 &  13.874 &  0.409 &  0.036 &  0.818 &   21.0      &  $-$1.3         &   5.6               &  $-$11.6          &   5.7  \\
BS~16545--0089  &  3.90 &  $-$3.49 &  14.450 &  0.318 &  0.023 &  1.356 & $-$161.5      &  14.0         &   2.0               &  $-$22.0          &   3.0  \\
CS~22948--0093  &  4.40 &  $-$3.43 &  15.180 &  0.360 &  0.013 &  1.346 &  369.2      & $-$15.6         &   5.1               &  $-$22.8          &   4.8  \\
CS~22965--0054  &  3.90 &  $-$2.84 &  15.069 &  0.497 &  0.110 &  1.532 & $-$281.7      &  26.0         &   1.0               &    0.0          &   3.0  \\
HE~1148--0037   &  3.70 &  $-$3.46 &  13.614 &  0.404 &  0.022 &  0.844 &  $-$10.9      & $-$46.0         &   8.0               &  $-$42.0          &   1.0  \\
{\CD}    &  4.40 &  $-$3.40 &  12.180 &  0.330 &  0.027 &  0.386 &  136.6      & 202.0         &   2.3               & $-$185.8          &   1.6  \\
SDSS~0040+16   &  4.40 &  $-$3.29 &  15.231 &  0.310 &  0.052 &  1.714 &  $-$49.4      &  24.0         &   4.0               &  $-$20.0          &   2.0  \\
SDSS~1033+40   &  4.40 &  $-$3.24 &  15.990 &  0.360 &  0.015 &  2.150 & $-$132.9      &   7.7         &   6.2               &   14.7          &   6.5  \\
{\Ga}        &  4.40 &  $-$3.37 &  11.453 &  0.385 &  0.003 &  0.217 &  434.7      &$-$230.4         &   2.4               &  $-$80.3          &   1.8  \\
{\Gb}        &  4.40 &  $-$3.23 &  11.140 &  0.370 &  0.014 &  0.210 &   76.7      & $-$54.3         &   3.6               & $-$399.5          &   1.8  \\
HD~84937       &  3.90 &  $-$2.15 &   8.280 &  0.410 &  0.007 &  0.073 &  $-$14.6      & 373.8         &   1.2               & $-$774.7          &   0.4  \\
{\BD}     &  3.90 &  $-$2.28 &   9.370 &  0.350 &  0.010 &  0.137 &  -91.2      &   0.9         &   0.6               & $-$172.4          &   1.4  \\
\tableline
\enddata
\end{deluxetable}

\begin{deluxetable}{llcrrrrcrrr}
\tablewidth{0pt}
\tabletypesize{\footnotesize}
\tablecaption{\label{tab:kine2}$A$(Li) and Derived Kinematic Parameters}
\tablehead{
  Star Name   & Pop\tablenotemark{a} &     $A$(Li) &  $U$        &  $V$        &    $W$      & $V_{\phi}$  &  ecc  & $r_{\rm max}$ & $r_{\rm min}$ & $Z_{\rm max}$ \\
              &                      &           &(km~s$^{-1}$)&(km~s$^{-1}$)&(km~s$^{-1}$)&(km~s$^{-1}$)&       &   (kpc)   &   (kpc)   &   (kpc) \\
}
\startdata
BS~16023--0046 & IH                   &  2.18     &     9       &  $-$237       &    30       &  $-$17        & 0.901 &  8.3      &  0.4      &  1.1    \\
BS~16968--0061 & IH                   &  2.17     &    51       &  $-$165       &   $-$44       &   54        & 0.722 &  8.3      &  1.3      &  1.1    \\
BS~17570--0063 & IH/OH                &  2.05     &    60       &  $-$322       &   $-$45       & $-$108        & 0.499 &  9.3      &  3.1      &  1.3    \\
CS~22177--0009 & IH/OH                &  2.20     &  $-$298       &   $-$80       &   147       &  126        & 0.840 & 32.8      &  2.8      & 13.9    \\
CS~22888--0031 & IH/OH                &  2.03     &   196       &  $-$153       &    50       &   65        & 0.788 & 12.5      &  1.4      &  2.8    \\
CS~22948--0093 & IH/OH                &  1.92     &  $-$329       &  $-$120       &  $-$201       &   99        & 0.854 & 38.3      &  3.0      & 30.9    \\
CS~22953--0037 & FI                   &  2.16     &   191       &   109       &    89       &  338        & 0.677 & 37.2      &  7.1      &  8.0    \\
CS~22965--0054 & IH                   &  2.06     &   250       &  $-$189       &    90       &    0        & 0.999 & 16.3      &  0.0      &  7.5    \\
CS~22966--0011 & IH                   &  1.91     &   $-$19       &  $-$164       &     7       &   55        & 0.716 &  8.3      &  1.3      &  0.8    \\
CS~29499--0060 & FI                   &  2.16     &    82       &    51       &    53       &  270        & 0.346 & 15.7      &  7.6      &  2.3    \\
CS~29506--0007 & IH                   &  2.15     &  $-$106       &  $-$126       &   $-$56       &   98        & 0.557 &  8.8      &  2.5      &  1.8    \\
CS~29506--0090 & IH                   &  2.10     &   $-$90       &  $-$172       &    24       &   51        & 0.758 &  8.6      &  1.1      &  0.8    \\
CS~29518--0020 & TD/IH                &  2.13     &    43       &   $-$68       &    38       &  152        & 0.304 &  8.8      &  4.7      &  1.2    \\
CS~29518--0043 & IH                   &  2.14     &    85       &  $-$105       &  $-$121       &  116        & 0.453 & 10.3      &  3.8      &  4.6    \\
CS~29527--0015 & IH                   &  2.08     &   150       &  $-$209       &   $-$81       &    7        & 0.970 & 11.3      &  0.1      &  2.9    \\
CS~30301--0024 & TD/IH                &  2.10     &    60       &  $-$108       &   $-$23       &  111        & 0.493 &  8.4      &  2.8      &  0.6    \\
CS~30339--0069 & IH                   &  2.13     &    68       &  $-$176       &   $-$22       &   44        & 0.774 &  8.8      &  1.1      &  1.6    \\
CS~31061--0032 & TD                   &  2.22     &   $-$19       &   $-$15       &   $-$33       &  205        & 0.073 &  9.1      &  7.9      &  0.9    \\
BS~16545--0089 & IH                   &  2.02     &  $-$192       &   $-$87       &   $-$95       &  132        & 0.655 & 16.3      &  3.4      &  4.4    \\
CS~22948--0093 & OH                   &  1.96     &  $-$332       &  $-$120       &  $-$205       &   99        & 0.857 & 39.8      &  3.0      & 32.4    \\
CS~22965--0054 & IH/OH                &  2.16     &   250       &  $-$190       &    90       &    0        & 1.000 & 16.3      &  0.0      &  7.5    \\
HE~1148--0037  & IH                   &  1.90     &    71       &  $-$185       &  $-$124       &   39        & 0.727 & 9.29      &  1.47     &  5.8    \\
{\CD}   & OH                   &  2.08     &   135       &  $-$388       &  $-$294       & $-$169        & 0.607 & 28.9      &  7.0      & 23.7    \\
SDSS~0040+16  & IH/OH                &  1.99     &    91       &  $-$269       &  $-$110       &  $-$59        & 0.688 & 10.6      &  1.9      &  5.0    \\
SDSS~1033+40  & FI                   &  2.09     &   $-$86       &   127       &  $-$101       &  347        & 0.602 & 37.1      &  9.2      &  9.8    \\
{\Ga}       & OH                   &  2.18     &   $-$48       &  $-$308       &   391       &  $-$89        & 0.657 & 40.2      &  8.3      & 39.0    \\
{\Gb}       & IH/OH                &  2.04     &  $-$183       &  $-$332       &  $-$131       & $-$113        & 0.642 & 14.3      &  3.1      &  6.1    \\
HD~84937      & IH                   &  2.26     &  $-$214       &  $-$203       &    0       &   15        & 0.955 & 13.5      &  0.3      &  0.0    \\
{\BD}    & TD/IH                &  2.28     &   $-$51       &  $-$112       &   $-$52       &  108        & 0.497 &  8.7      &  2.9      &  0.9    \\

\tableline
\enddata
\tablenotetext{a}{Assigned Stellar Population -- TD: Thick Disk (including Metal-Weak Thick Disk); IH: Inner Halo; OH: Outer Halo; FI: Further Inspection required}
\end{deluxetable}

\end{document}